\documentclass[12pt,preprint]{emulateapj}

\usepackage{graphicx,amsmath,natbib,color}

\usepackage{natbib}
\usepackage{color}
\usepackage{amsmath,amsthm,amssymb}
\usepackage{epsfig}
\usepackage{subfigure}
\usepackage{mathrsfs}
\usepackage{url}

\slugcomment{Appected to ApJ Letters. Last Compiled \today}

\shorttitle{Early Results from VLT SPHERE: Spectroscopy of 2M0122B }
\shortauthors{Hinkley et al.}

\begin{document}

\title{
Early Results from VLT SPHERE: Long-Slit Spectroscopy of 2MASS 0122-2439\,B, a young companion near the Deuterium Burning Limit.\footnotemark[\lowercase{a}]}

\author{Sasha Hinkley\altaffilmark{1}}
\author{Brendan P. Bowler\altaffilmark{2,8}}
\author{Arthur Vigan\altaffilmark{3,4}}
\author{Kimberly M. Aller\altaffilmark{5}}
\author{Michael C. Liu\altaffilmark{5}}
\author{Dimitri Mawet\altaffilmark{6,4}}
\author{Elisabeth Matthews\altaffilmark{1}}
\author{Zahed Wahhaj\altaffilmark{4}}
\author{Stefan Kraus\altaffilmark{1}}
\author{Isabelle Baraffe\altaffilmark{1,7}} 
\author{Gilles Chabrier\altaffilmark{1,7}}

\altaffiltext{1}{University of Exeter, Astrophysics Group, Physics Building, Stocker Road, Exeter, EX4 4QL, UK. }
\altaffiltext{2}{Division of Geological and Planetary Sciences, California Institute of Technology, 1200 E. California Blvd, Pasadena, CA 91125, USA.}
\altaffiltext{3}{Aix Marseille Universit\'e, CNRS, LAM (Laboratoire d`Astrophysique de Marseille) UMR 7326, 13388 Marseille, France.}
\altaffiltext{4}{European Southern Observatory, Alonso de Cordova 3107, Vitacura, Santiago, Chile}
\altaffiltext{5}{University of Hawaii, Institute of Astronomy, 2860
  Woodlawn Drive, Honolulu, HI 96822, USA }
\altaffiltext{6}{Department of Astronomy, California Institute of Technology, Mail Code 249-17, 1200 E. California Blvd, Pasadena, CA 91125, USA}
\altaffiltext{7}{CRAL, ENS-Lyon (CNRS UMR 5574), Lyon, France}
\altaffiltext{8}{Caltech Joint Center for Planetary Astronomy Fellow.}

\footnotetext[a]{Based on observations made with ESO Telescopes at the La Silla Paranal Observatory Under Program ID 060.A-9381.}

\begin{abstract}
We present 0.95--1.80 $\mu$m spectroscopy of the $\sim$12--27 $M_{\rm Jup}$ companion orbiting the faint ($R$$\sim$13.6), young ($\sim$120\,Myr) M-dwarf 2MASS J01225093--2439505 (``2M0122--2439\,B'') at 1$\farcs$5 separation (50\,AU).  Our coronagraphic long-slit spectroscopy was obtained with the new high contrast imaging platform VLT-SPHERE during Science Verification. The unique long-slit capability of SPHERE enables spectral resolution an order of magnitude higher than other extreme AO exoplanet imaging instruments. With a low  mass, cool temperature, and very red colors, 2M0122--2439 B occupies a particularly important region of the substellar color-magnitude diagram by bridging the warm directly imaged hot planets with late-M/early-L spectral types (e.g.~$\beta$ Pic b and ROXs 42Bb) and the cooler, dusty objects near the L/T transition (HR\,8799bcde and 2MASS 1207b). We fit BT-Settl atmospheric models to our $R$$\approx$350 spectrum and find $T_{\rm eff}$=1600$\pm$100~K and $\log(g)$=4.5$\pm$0.5 dex. Visual analysis of our 2M0122--2439 B spectrum suggests a spectral type L3-L4, and we resolve shallow $J$-band alkali lines, confirming its low gravity and youth.  Specifically, we use the \citet{al13} spectral indices to quantitatively measure the strength of the FeH, VO, KI, spectral features, as well as the overall $H$-band shape.  Using these indices, along with the visual spectral type analysis, we classify 2M0122--2439 B as an intermediate gravity ({\sc int-g}) object with spectral type L3.7$\pm$1.0. 

\end{abstract}




\section{Introduction}
The study of objects with masses and temperatures near the transition between the L and T spectral types, the so-called ``L/T transition objects,'' has emerged as one of the most interesting and perplexing phenomena of substellar physics \citep{bmk11,b14}.  Indeed, ongoing observations of substellar objects continue to confirm early indications that spectral types for L/T transition objects have a very strong dependence on surface gravity \citep[e.g.][]{mh06,lmd13}.  Specifically, for a given effective temperature, low-gravity objects tend to be classified with spectral types earlier than their effective temperatures would naively predict based on older, higher-gravity objects.  The effect of dust in the atmospheres of these young, low-gravity objects is very likely linked to this phenomenon \citep[e.g.][]{cba00, sm08, msc12}. However, the small number of {\it young} ($\lesssim$100-200\,Myr) objects near the L/T transition is currently preventing a detailed understanding of this process.   

Furthermore, even for modern atmospheric models, relatively narrow wavelength coverage for the small number of L/T transition objects has led to strong ambiguities in interpretation.   Only when data covering multiple, simultaneous bandpasses are obtained can strong constraints be placed on the physical processes present \citep[e.g.][]{she12,ims14}.   As an added limitation, much of our knowledge of these low-gravity phenomena in young L-dwarfs comes from a handful of objects: e.g.~ the directly-imaged planets HR\,8799bcde, $\beta$ Pic\,b, and 2M1207\,b  with {\it very} young ages of $\sim$10-30 Myr \citep[e.g.][]{zrs11,mb14}, as well as a handful of brown dwarf and planetary mass objects e.g. G196-3B \citep{rzm98}, HD203030\,B \citep{mh06}, 2MASS\,0141 \citep{kbb06}, PSO\,318-22 \citep{lda13, lmd13}, and WISE\,0047 \citep{gal15}.  A larger sample of objects spanning a wider range of ages is needed to fully disentangle the effects of age (and hence surface gravity) on the spectra of these objects.




\begin{table}
\caption{Spectral Type and Gravity Indices.}
\label{tab:sptgravity}
  \centering
  \begin{tabular}{ll}
  \hline
{\bf  Spectral Type Indices}\tablenotemark{a}             &                       \\    
Visual SpT ($J$-band)                   &  L4$\pm$1           \\
Visual SpT ($K$-band)                   &  L3$\pm$1           \\
H$_{2}$O                                &  \nodata              \\
H$_{2}$O--D                             &  \nodata              \\
H$_{2}$O--1                             &  14.1$_{-1.1}^{+1.1}$  \\
H$_{2}$O--2                             &  \nodata               \\ 
Final Index-based SpT                   &  14.1$_{-1.1}^{+1.1}$  \\ 
Final SpT                               &  L3.7$\pm$1            \\       
\hline 
{\bf Gravity Indices \& Classification}\tablenotemark{b} &                         \\ 
FeH$_{z}$                               & 1.13$_{-0.03}^{+0.03}$  \\
VO$_{z}$                                & 1.057$_{-0.011}^{+0.010}$  \\
KI$_{J}$                                & 1.095$_{-0.003}^{+0.003}$   \\
H-cont                                  & 0.942$_{-0.004}^{+0.004}$  \\
Overall Gravity Scores                  & 2011 (20??)                  \\
Gravity Class                           & {\sc int-g}                   \\
\hline 
\tablenotetext{1}{``Final index-based SpT'' is the median SpT calculated from the indices of the Monte Carlo realizations of the spectrum. ``Final SpT'' refers to the weighted mean of the visual SpT and the index-based SpT.}
\tablenotetext{2}{The gravity score in parenthesis refer to the \citet{al13} classification scheme. Objects with index values corresponding to {\sc int-g} but are within 1$\sigma$ of the {\sc fld-g} value are classified with a score of ``?''. 
}
\end{tabular}
\end{table}

\begin{figure*}
  \centering
  \includegraphics[width=1.0\textwidth]{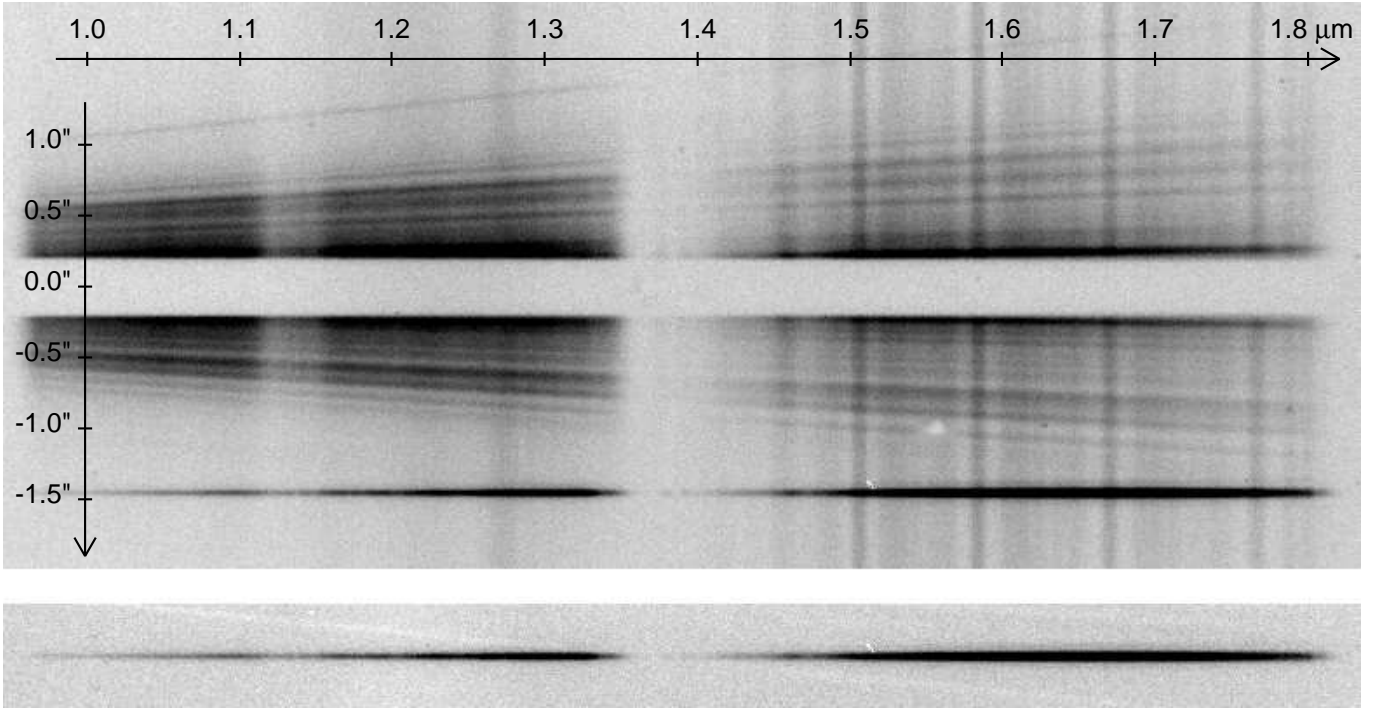}
  \caption{SPHERE-IRDIS medium resolution spectrum of 2M0122-2439 and its companion. The spectrum is dispersed in the horizontal direction, while the vertical direction indicates the spatial direction along the spectrograph slit.  The primary is hidden behind the 0.2\arcsec-radius coronagraphic mask (light horizontal band at the center), so only its halo and speckles are visible in the two-dimensional spectrum. The spectrum of the companion is clearly visible at 1.45\arcsec below the primary. Strong OH sky lines are also visible redward of 1.4~$\mu$m. The top panel shows the 2D spectrum of the primary and companion after combining all the integrations but before any subtraction of the stellar halo and speckles. 
The lower panel shows the spectrum of the companion after the subtraction detailed in \S\ref{sec:data_reduction_analysis}. 
}
  \label{fig:mrs_spectrum}
\end{figure*}

A mid-L companion has recently been discovered \citep{bls13} orbiting the young M3.5V star 2MASS J01225093-2439505 (hereafter ``2M0122-2439''), with an estimated mass of 12--27 M$_{\rm Jup}$. The young age of this system ($\sim$120 Myr), coupled with the intrinsic faintness of the host star means that even the relatively modest contrast observations of this companion ($\Delta H$=6.2) were sensitive to planetary mass companions ($\sim$5--20 M$_{\rm Jup}$).  This relatively nearby ($\sim$36 pc) host star, first recognized to possess strong X-ray emission by \citet{rgh06}, was subsequently assigned membership to the $\sim$120 Myr AB Dor young moving group \citep{mdl13}.  
Since much of our understanding of L-dwarf companions comes from systems with ages $\sim$10-30 Myr, the somewhat older age of the 2M0122-2439  system makes it particularly important for mapping out the evolution of giant planets and brown dwarfs.   Furthermore, its low temperature coupled with its very red colors make this companion a ``bridge'' between hot ($\sim$1700--2000\,K) directly imaged planets with late-M/early-L spectral type \citep[e.g.~ROXs 42Bb and $\beta$ Pic b,][]{rkl05, kic14,cdd14, cbf15} and cooler ($\sim$800--1300\,K), dusty planetary mass companions \citep[e.g.~HR\,8799bcde and 2M1207b,][]{cld04, mzk10}.

In this paper, we present near-infrared (NIR) spectroscopy of 2M0122-2439\,B using the newly commissioned Spectro-Polarimetric High-Contrast Exoplanet REsearch (SPHERE) high contrast imaging platform at VLT obtained during science verification observations. Our observations were obtained using a combination of coronagraphy and long-slit spectroscopy, unique to the SPHERE instrument.  In \S2 and 3, we describe the observations and analysis.  In \S4, we describe our results of spectral typing and constraints on the surface gravity measurements.

\begin{figure*}
\epsscale{0.95}
\vskip -1 in
\plotone{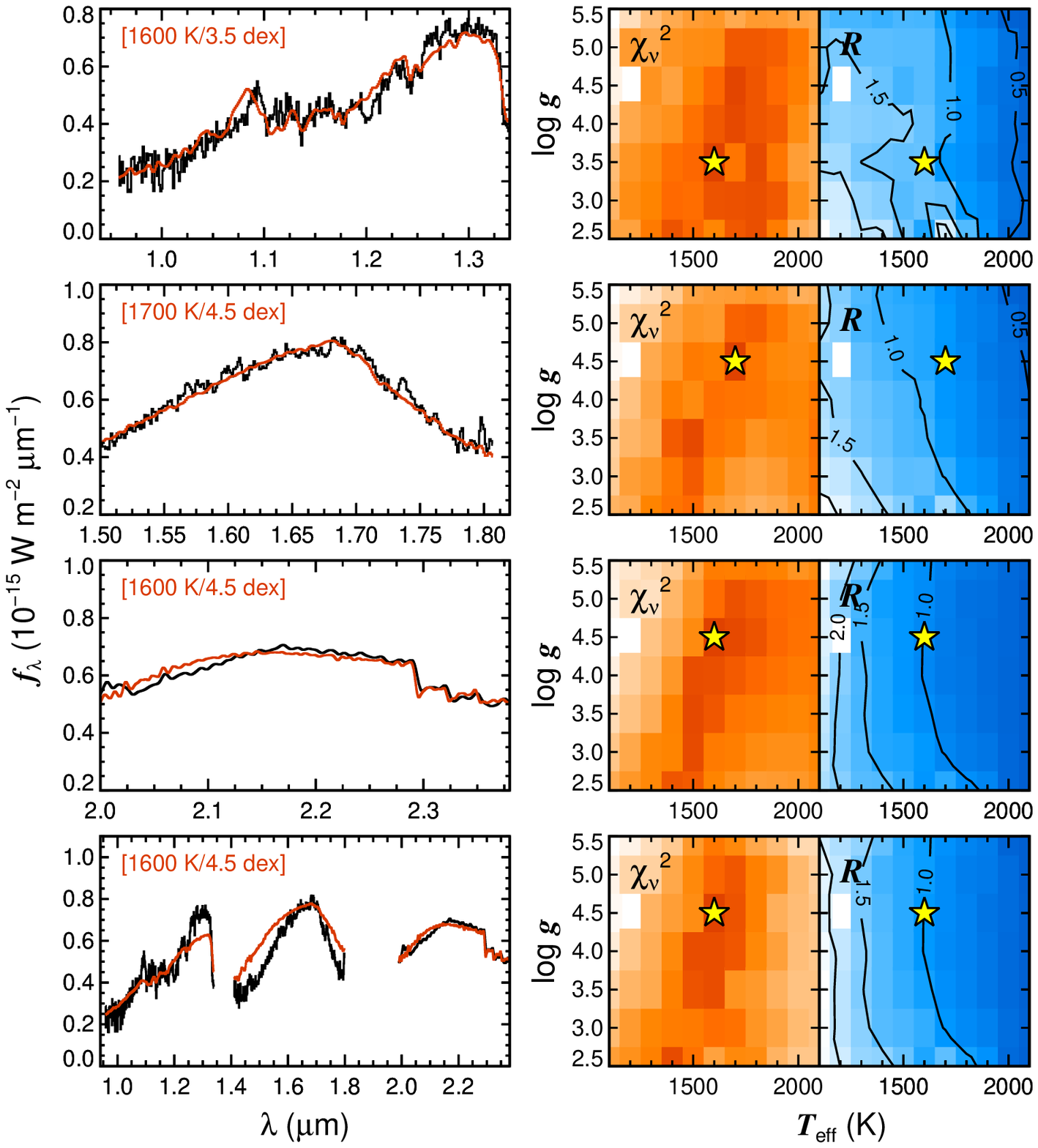}
\vskip -.1 in
\caption{Atmospheric model fits (red) to our new SPHERE 0.9--1.8~$\mu$m spectrum of 2M0122-2439~B together with the $K$-band OSIRIS spectrum from \citet[black]{bls13}.  The BT-Settl models from \citet{ahf11} yield temperatures of 1600--1700~K and a surface gravity of 3.5-4.5~dex among individual band fits and the entire 0.9--2.4~$\mu$m spectrum.  The right panels display reduced $\chi^2$ and inferred radius contours (in $R_\mathrm{Jup}$) assuming a distance of 36~$\pm$~4~pc. 
}. 
\label{modfit}
\end{figure*}

\section{Observations}
Our observations were gathered on UT 06 December 2014 during the Science Verification period of the SPHERE instrument \citep[e.g.][]{bfd08} at the Very Large Telescope (VLT) located at Cerro Paranal in Chile. The data were obtained using the Infrared Dual-band Imager and Spectrograph (IRDIS; \citealt{dohlen2008}) instrument configured in the long slit spectroscopy (hereafter ``LSS'') mode, using the 0.09'' slit providing medium resolution ($R$$\approx$350) across the $YJH$-bands \citep{vigan2008}. This observing mode is unique to SPHERE with the coronagraphic mask embedded into the slit to create a long slit coronagraph, enabling a spectral resolution an order of magnitude higher than the integral field spectrographs found in SPHERE and other new high-contrast instruments \citep[e.g.][]{hoz11, mgi14}.

The observations consisted of 55 coronagraphic observations 
resulting in a total exposure time of 59 minutes on target.  
Subsequent to this, a 20s observation of HD\, 10342, an A0V spectroscopic standard placed off the coronagraphic mask, but within the slit, was obtained. 
No sky backgrounds were acquired. Despite the faintness of the host star ($R$=13.6), Strehl ratios of $\sim$70\% were achieved throughout the observations thanks to the SAXO extreme adaptive optics system \citep{fusco2013}. The DIMM seeing was stable around 1.0\arcsec~throughout the first half of the observations, and no seeing data were available for the second half.
Wavelength calibration was acquired during the night at the end of the science observations using arc lamps.  


\section{Data Reduction \& Analysis}
\label{sec:data_reduction_analysis}
Standard calibrations were produced using the preliminary release (v0.14.0-2) of the SPHERE Data Reduction and Handling (``DRH'') pipeline \citep{pavlov2008}. 
The 55 science frames were median combined, producing a two-dimensional spectrum that is shown in the top panel of Figure~\ref{fig:mrs_spectrum}. In addition to the strong signal from the companion, residual, uncorrected quasi-static speckles \citep[e.g.][]{hos07,cpb11} can be seen tracing diagonal paths with wavelength. 

For the subtraction of the stellar halo and speckles at the position of the companion, we tried two approaches, both providing similar results. The first approach is similar to spectral differential imaging adapted for LSS data, as presented in \citet{vigan2008} and demonstrated on-sky in \citet{vigan2012}. The second approach is much simpler as it does not involve any assumption on the spectral dependence of the diffraction features: at the position of the companion, we simply subtracted the diametrically symmetric value of the halo with respect to the star. The validity of this latter approach owes to the faintness of the primary target and companion, which results in the companion being in an area of the spectrum where the data are limited by the photon noise of the halo. 
Both attempted halo subtraction schemes provide a very good subtraction leaving negligible spectroscopic imprint. The bottom panel of Figure~\ref{fig:mrs_spectrum} shows the results of this exercise.

The spectrum of the companion and the spectroscopic standard were extracted in the following way: an aperture centered at the object position was created with a width of $\epsilon~\lambda/D$ in each spectral channel and the signal was then summed in each channel to obtain a one-dimensional spectrum. The value of $\epsilon$ was varied from 0.5 to 1.5, but no differences were observed in the output spectrum as a function of the aperture width. 
The only difference between the extraction of the spectrum of the companion and the spectroscopic standard is the position of the center of the mask in the data. The local noise was estimated by summing the residuals after subtraction on either side of the companion spectrum using the same aperture size as the companion. We verified that the spectrum extracted with the two subtraction schemes and the different aperture sizes lie within these error bars, making us confident that our extraction of the spectrum is correct.  

Finally, a telluric correction on the extracted spectrum was performed using the ``xtellcor\_general'' software package \citep[part of the IRTF SpeXtool reduction package;][]{vcr03, cvr04}, using the extracted spectrum of the spectroscopic standard star.  
Despite these wavelength calibration tests, a slight, but uniformly systematic wavelength offset (0.0025$\mu$m) persisted in the redward direction.  To compensate for this, we manually shifted our spectra by this amount.  This offset was determined by performing a cross correlation with the 2M0122-2439\,B spectra with an L6 brown dwarf obtained from the IRTF library.  
Such offsets in the final reduced spectra can naturally arise due to non-perfect centering of the faint companion point spread function in the slit.  
Comparison of the $H$-band spectrum from Keck/OSIRIS \citep{bls13} with the $H$-band spectrum from SPHERE shows excellent agreement.

\begin{figure*}
  \begin{center}
    \includegraphics[width=0.75\textwidth,angle=90]{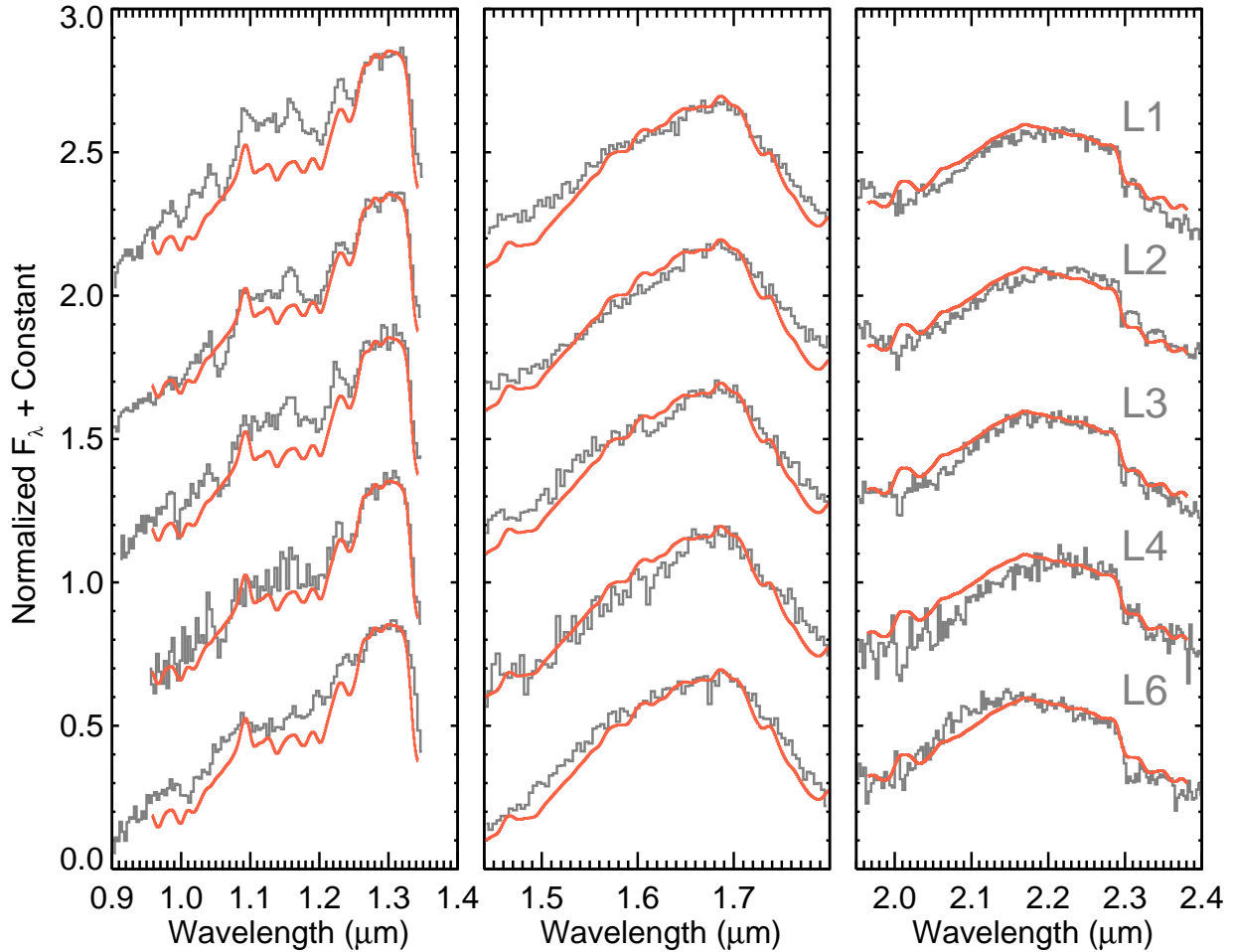}
    \caption{A visual spectral type classification of 2M0122$-$2439\,B
      (\emph{dark orange}) smoothed to $R$$\approx$130 (to match the
      resolution of our comparison spectra) compared with a collection of L~dwarf {\sc vl-g} and {\sc int-g} objects taken from \citet{al13}. 
      The standards are, from top to bottom:
      2M0117-34 (L1), 2M0536$-$19 (L2), 2M1726+25 (L3), 2M1551$+$09 (L4), and 2M0103+19 (L6).
      We classify
      2M0122$-$2439\,B as an L4$\pm$1 in the J-band and L3$\pm$1 in the $K$-band using
      this visual comparison. 
      \label{fig:visualclass}}
  \end{center}
\end{figure*}

\section{Results}

\subsection{Physical Properties}

We performed model fits to our VLT SPHERE 0.9--1.8~$\mu$m spectrum together with the Keck/OSIRIS $K$-band spectrum from \citet{bls13} to infer the physical properties of 2M0122-2439\,B.  We fit the solar metallicity ``CIFIST2011bc'' version of the BT-Settl models of \citet{ahf11} with no $\alpha$-element enhancement to individual $Y+J$ (0.9--1.35~$\mu$m), $H$ (1.5--1.8~$\mu$m), and $K$ (2.0--2.4~$\mu$m) bands as well as the full 0.9--2.4~$\mu$m spectrum.  The models are first Gaussian smoothed to the same resolution as the SPHERE data ($R$$\approx$ 350) and interpolated onto the wavelength sampling of our spectrum.  Our analysis is carried out in a Monte Carlo fashion incorporating spectral measurement errors and band-to-band flux calibration uncertainties as described in \citet{blk11}. Briefly, for each trial we generate a new spectrum by randomly adding Gaussian noise drawn from spectral errors at each pixel.  Likewise, a new flux calibration scale factor is drawn from a Gaussian distribution for each trial based on the measured photometry and uncertainties of 2M0122-2439\,B.  Note that only two flux calibration scale factors are considered here: one for our SPHERE 0.9--1.8~$\mu$m spectrum and one for the OSIRIS $K$-band spectrum.  Following Cushing et al. (2005) we compute a model scale factor, equal to $R^2$/$d^2$, and a reduced $\chi^2$ value for each BT-Settl model spanning $T_\mathrm{eff}$ = 1100--2100\,K ($\Delta$ $T_\mathrm{eff}$=100\,K) and $\log(g)$ = 2.5--5.5~dex ($\Delta$log$(g)$ = 0.5~dex).

\begin{figure*}
  \begin{center}
    \includegraphics[width=0.80\textwidth,angle=0]{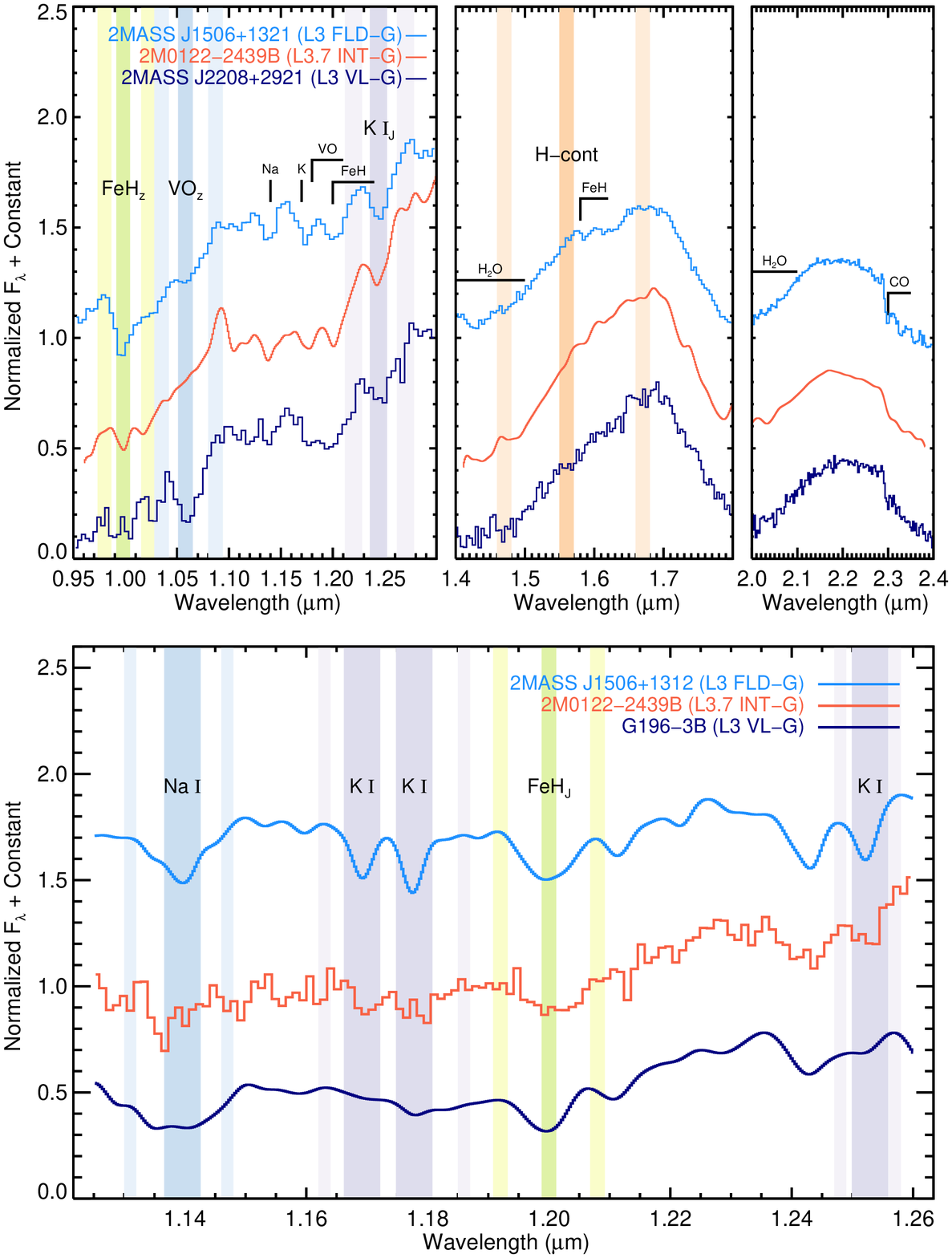}
    \caption{{\it Top panels:} NIR spectrum from VLT/SPHERE ($YJH$-bands) and Keck/OSIRIS
      \citep[$K$ band,][]{bls13} of 2M0122$-$2439\,B (\emph{red}), compared with a
      field/{\sc fld-g} (\emph{light blue}) and young/{\sc vl-g}
      (\emph{dark blue}) spectral standard of similar spectral
      type. For comparison spectra, we used the field 
      standard ({\sc fld-g}) and the very low gravity ({\sc vl-g}) spectral standards from 
      \citet{klb10} and \citet{al13}, respectively. In the top panels, we have smoothed our spectrum of
      2M0122$-$2439\,B to $R$$\approx$130 to match the resolution of the
      comparison spectra. Gravity-sensitive features from
      \citet{al13} are labeled and the wavelength
      ranges used to calculate gravity indices are highlighted for
      FeH$_{z}$ (\emph{yellow-green}), VO$_{z}$ (\emph{blue}),
      KI$_{J}$ (\emph{purple}), and H-cont
      (\emph{orange}). 
      {\it Lower panel:} the J-band spectrum in its native resolution ($R$$\approx$350), along with the well-characterized {\sc vl-g} dwarf G196-3B \citep{rzm98} and the same field object as the top panels. 
      The comparison spectra have been smoothed to the
      resolution of the SPHERE spectrum. Gravity-sensitive features in the $J$-band from
      \citet{al13} are labeled and the wavelength
      ranges used to calculate gravity indices are highlighted for
      Na~I (\emph{blue}), K~I (\emph{purple}), and FeH$_{J}$
      (\emph{yellow-green}), though these are for illustration
      purposes only, since our spectrum of 2M0122$-$2439\,B is too low
      resolution to use these gravity indices. 
      \label{fig:sxdind}
      }
  \end{center}
\end{figure*}

Figure~\ref{modfit} shows the best-fit models, the median reduced $\chi^2$ contour maps from 1000 Monte Carlo trials, and the median inferred radius contour map based on the model scale factor and distance estimate of 36~$\pm$~4~pc.  The [1600~K/4.5~dex] model provides the best-fit to the full 0.9--2.4~$\mu$m spectral regions and yields a radius of 1.0\,$R_\mathrm{Jup}$.  The $H$-band yields a slightly warmer model (1700~K/4.5~dex) and smaller radius (0.8\,$R_\mathrm{Jup}$), while the $YJ$-band fit yields a slightly lower gravity model ($\log(g)$=3.5~dex) with a slightly larger radius (1.2\,$R_\mathrm{Jup}$).  The Monte Carlo procedure produces uncertainties in the temperature and gravity of 100~K and 0.5 dex, respectively.  These best fit temperature values are somewhat warmer than the effective temperature predicted by evolutionary models, which ranges from 1350--1500~K (\citealt{bls13}). Possible reasons for this discrepancy could be imperfections in opacity sources, line lists, and treatment of dust.  Also, as \citet{bls13} emphasize, this target is unusually red, possibly exacerbating the 0.95-2.35 $\mu$m fit. It is also worth noting that the covariance in the $\log(g)/T_{\rm eff}$ planes shown in Figure~\ref{modfit} also have low $\chi^2$ values.


\subsection{Spectral Type \& Luminosity}

We determined the NIR spectral type of 2M0122$-$2439\,B using both the
index-based (``quantitative'') and visual methods described in \citet{al13}. For
consistency with the gravity analysis (\S4.3), we used the spectrum smoothed to the same resolution
($R$$\approx$150) for the index-based spectral typing. The index-based
method combines the spectral-type sensitive indices from
\citet{ajl07}, \citet{Slesnick04}, and
\citet{mmb03} to calculate the spectral type. In our
case, the H$_{2}$O-D \citep{mmb03} index is not
applicable because our spectrum does not include the wavelengths used
for the index. Also, the H$_{2}$O \citep{ajl07} index is
not valid because the spectral types calculated for this index ($>$L4) are always out of the valid
range (M5--L4). 
Although all the other indices are valid for L~dwarfs,
they are not valid across the entire spectral class. The H$_{2}$O-1 and
H$_{2}$O-2 \citep{Slesnick04} indices are valid for spectral
types of M4--L5 and M4--L2, respectively. As such, only the H$_{2}$O-1 index is applicable for this object, leading to an index-based SpT of 14.1$\pm$1.1 (See Table~\ref{tab:sptgravity}).

Secondly, we also visually compared 2M0122$-$2439\,B to L~dwarf
low-gravity ({\sc vl-g}) and ({\sc int-g}) spectroscopic standards defined in \citet{al13}. We used standard spectra taken from the IRTF Spectral Library
\citep{crv05} and the SpeX Prism Library (\url{http://pono.ucsd.edu/$\sim$adam/browndwarfs/spexprism}).
Following the visual classification methods for young and intermediate
age objects of \citet{al13}, we normalize both our
candidates and the standard template in each NIR band separately (see
Figure~\ref{fig:visualclass} for an example). We also smoothed our
spectrum to R$\approx$130 in order to compare with the publicly available
L~dwarf spectroscopic standards, which were taken at a similar
resolution. 
Since the sequence presented in Figure~\ref{fig:visualclass} is a mixture of {\sc vl-g} and {\sc int-g} gravity classes, it should not be treated as a proper Spectral Type sequence, but is still useful for comparison. 
We assumed an uncertainty of $\pm$1 subtype for our visual classification in
both the $J$ and $K$-band \citep[consistent with][]{al13}. 
Using this visual comparison method, we classify 2M0122$-$2439\,B as L4.0$\pm$1.0 in the $J$-band and
L3.0$\pm$1.0 in the $K$-band. The final spectral type is calculated following
\citet{al13} and is the weighted mean of each of
the visual and index-based spectral types, with an adopted uncertainty
of 1 subtype. This gives a final spectral type of L3.7$\pm$1.0.  

A bolometric luminosity of 2M0122-2439 B is calculated by integrating the flux-calibrated model spectrum derived in \S4.1 between wavelengths of 0.25 and 1000~$\mu$m. This gives a value of $\log(L/L_\odot$)=$-4.16\pm$0.10 dex with the uncertainty dominated by uncertainty in distance determination. This is in good agreement with the $\log(L/L_\odot$)=$-4.19\pm$0.10 dex cited in \citet{bls13}.

\subsection{Gravity Classification}
Young ultracool dwarfs have lower surface gravity than field objects, resulting in distinctive features in their NIR spectra \citep[e.g.][]{ajl07, al13}. 
Low resolution ($R$$\approx$35) spectroscopy \citep[e.g.][]{hob08} can place strong constraints on the effective temperature, but constraints on surface gravity are more challenging \citep[e.g.][]{hpf13}, requiring higher resolution. 
At higher spectral resolution, young objects tend to have weaker FeH and alkali line absorption and stronger VO absorption in the $J$-band, in addition to a triangular-shaped $H$-band (See Figure~\ref{fig:sxdind}, upper panels). 
We compared these spectral features of 2M0122$-$2439\,B with known old, field gravity ({\sc fld-g}) and young, very low gravity ({\sc vl-g}) dwarfs of the same spectral type to visually assess the surface gravity (Figure~\ref{fig:sxdind}, upper panels). For this visual comparison, we smoothed the spectrum of 2M0122$-$2439\,B to $R$$\approx$130 to match the native resolution of our comparison spectra. 2M0122$-$2439\,B has weak FeH$_{z}$, slightly weak KI$_{J}$ absorption, and a triangular-shaped $H$-band, all signatures of low gravity. However, the VO$_{z}$ absorption is weak, which is to be expected for this SpT \citep[][specifically their Figure 20]{al13}. We also examined the alkali lines in the $J$~band, which are resolved at the native spectral resolution of $R$$\approx$350 (Figure~\ref{fig:sxdind}, lower panel). 2M0122$-$2439\,B may have slightly weak KI~[1.253~$\mu$m] and FeH$_{J}$ lines similar to the {\sc vl-g} comparison spectrum. However, the NaI~[1.140~$\mu$m], KI~[1.169~$\mu$m] and KI~[1.177~$\mu$m] lines have too low signal-to-noise to place meaningful constraints on the surface gravity. Overall, the spectral features of 2M0122$-$2439\,B suggest that it has a low surface gravity. 

We then quantitatively assessed the gravity classification of our objects using spectral indices from \citet{al13}. Under this classification scheme, indices are measured in the $J$ and $H$-bands and are each then assigned a score (0, 1, or 2) according to the index value and the spectral type of the object, with higher numbers indicating lower gravity. However, index values corresponding to {\sc int-g} but within 1$\sigma$ of the {\sc fld-g} value are classified with a score of ``?'', instead of 1, and ignored. Indices that are not valid because of either the spectral type and/or the resolution of an object are scored as ``n''. These scores are combined into a final 4-number gravity score that represents the FeH, VO, alkali lines, and H-band continuum indices (e.g. 0110, 2110, etc.). Finally, this gravity score is used to determine the overall gravity classification for the object: field gravity ({\sc fld-g}), intermediate gravity ({\sc int-g}), or very low gravity ({\sc vl-g}).

We also used a modified classification method (Aller et
al., submitted), which uses a Monte Carlo simulation to propagate
measurement uncertainties into the overall gravity classification
uncertainties. However, unlike \citet{al13} this method does not
give a score of ``?'' for indices with values corresponding to {\sc
  int-g} but within 1$\sigma$ of the {\sc fld-g} values. Instead, this method
gives them a score of 1 and thus does not ignore the score. The overall
gravity classification value is the median from the values calculated
in the Monte Carlo realizations. This allows us to better flag
borderline objects between {\sc int-g} and {\sc fld-g}.

Our spectrum of 2M0122$-$2439\,B is too low resolution ($R$$\approx$350) to
use the alkali line indices from \citet{al13}
(e.g. Figure~\ref{fig:sxdind}, lower panel) in the $J$ band which require at least
$R$$\approx$750. Therefore, we assessed the gravity classification
of 2M0122$-$2439\,B using the \citet{al13} low resolution
gravity indices: FeH$_{z}$, VO$_{z}$, KI$_{J}$, and H-cont (Table~\ref{tab:sptgravity}). To
calculate these indices, we followed \citet{al13} and
smoothed the spectrum to $R$$\approx$150, because the indices are tailored
for that resolution. Our final gravity classification of
2M0122$-$2439\,B is {\sc int-g}.

\section{Summary \& Conclusions}
Using a novel combination of long-slit spectroscopy and coronagraphy with the VLT-SPHERE instrument, we have obtained $R$$\approx$350 spectra of the young, early/mid L-dwarf 2M0122-2439\,B.  Atmospheric model fits to our spectra suggest a surface gravity and temperature $T_{\rm eff}$=1600$\pm$100~K  and $\log(g)$=4.5$\pm$0.5 dex. We also resolve shallow $J$-band alkali lines in 2M0122--2439 B, and use the spectral indices defined in \citet{al13} to measure the strength of the FeH, VO, KI, spectral features, as well as the overall $H$-band shape.  Using these indices, we confirm the low gravity and youth of 2M0122--2439\,B. Visual classification alone suggests a spectral type L3-L4, and the index-based classification scheme outlined in \citet{al13}, suggests a L4.1$\pm$1.1 spectral type. Combining both of these methods yields a final spectral type of L3.7$\pm$1.



\acknowledgments
We thank the anonymous referee for several useful coments. The research of K.M.A. was supported by the National Science Foundation Graduate Research Fellowship under Grant No. DGE-1329626. Any opinion, findings, and conclusions or recommendations expressed in this material are those of the authors' and do not necessarily reflect the views of the National Science Foundation. S.K. acknowledges support from an STFC Ernest Rutherford Fellowship ST/J004030/1, Ernest Rutherford Grant (ST/K003445/1), and Marie Curie CIG grant (SH-06192). This work is partly supported by Royal Society award WM090065 and the consolidated STFC grant ST/J001627/1.  The research leading to these results has received funding from the European Research Council under the European CommunityÕs Seventh Framework Programme (FP7/2007-2013 Grant Agreement no. 247060).


\end{document}